\documentclass[twocolumn,preprintnumbers,showpacs,pre,floatfix,amsmath,amssymb,notitlepage]{revtex4-1}
\usepackage{epsfig}
\usepackage{subfigure}
\usepackage{amsmath}
\usepackage{color}
\usepackage{amssymb}
\usepackage{setspace}
\usepackage{graphicx}% Include figure files
\usepackage{dcolumn}% Align table columns on decimal point
\usepackage{bm}% bold math
\usepackage{times}
\usepackage{enumerate}
\usepackage{footnote}
\usepackage{cancel}
\input epsf

\newtheorem{theorem}{Theorem}
\newtheorem{definition}[theorem]{Definition}

\newcommand{\appropto}{\mathrel{\vcenter{
  \offinterlineskip\halign{\hfil$##$\cr
    \propto\cr\noalign{\kern2pt}\sim\cr\noalign{\kern-2pt}}}}}

\graphicspath{{figures/}}

\begin{document}

\author{Per Sebastian Skardal}
\email{persebastian.skardal@trincoll.edu} 
\affiliation{Department of Mathematics, Trinity College, Hartford, CT 06106, USA}

\author{Kirsti Wash}
\email{kirsti.wash@trincoll.edu} 
\affiliation{Department of Mathematics, Trinity College, Hartford, CT 06106, USA}

\title{Spectral properties of the hierarchical product of graphs}

\begin{abstract}
The hierarchical product of two graphs represents a natural way to build a larger graph out of two smaller graphs with less regular and therefore more heterogeneous structure than the Cartesian product. Here we study the eigenvalue spectrum of the adjacency matrix of the hierarchical product of two graphs. Introducing a coupling parameter describing the relative contribution of each of the two smaller graphs, we perform an asymptotic analysis for the full spectrum of eigenvalues of the adjacency matrix of the hierarchical product. Specifically, we derive the exact limit points for each eigenvalue in the limits of small and large coupling, as well as the leading-order relaxation to these values in terms of the eigenvalues and eigenvectors of the two smaller graphs. Given its central roll in the structural and dynamical properties of networks, we study in detail the Perron-Frobenius, or largest, eigenvalue. Finally, as an example application we use our theory to predict the epidemic threshold of the Susceptible-Infected-Susceptible model on a hierarchical product of two graphs.
\end{abstract}


\maketitle

\section{Introduction}\label{sec1}

Graphs and networks represent fundamental structures that describe the patterns of interactions throughout nature and society~\cite{Newman}, examples of which include electrical power grids~\cite{Motter2013NatPhys}, faculty hiring networks~\cite{Clauset2015SciAdv}, protein-protein interaction networks~\cite{Tamames2007GenBio}, and the neurons in the brain~\cite{Larremore2011PRL}. Large graphs and networks are often comprised of several smaller pieces, for example motifs~\cite{Milo2002Science}, communities or modules~\cite{Girvan2002PNAS,Porter2006PNAS}, layers~\cite{DeDomenico2013PRX}, or self-similar subnetwork structures~\cite{Guimera2003PRE}. Moreover, the macroscopic properties of such large graphs are often determined by the agglomeration of properties of these smaller structures~\cite{Arenas2006PRL,Skardal2012PRE}. One natural way to construct a graph from two or more smaller graphs is by the well-known Cartesian product~\cite{Hammack2011}. Recently, Barri\`{e}re et al. introduced a generalization of the Cartesian product known as the hierarchical product~\cite{Barriere2009DAM1,Barriere2009DAM2}, which captures connectivity characteristics that are less regular and therefore more heterogeneous than those found in the Cartesian product.

A great deal of research has shown that both structural and dynamical properties of a given graph or network are determined by the eigenvalue sprectrum of its associated coupling matrices~\cite{Newman}. We consider here a graph's adjacency matrix: an $N\times N$ matrix $A$ whose entries correspond to edges such that $A_{ij}=1$ if vertices $i$ and $j$ are connected, and otherwise $A_{ij}=0$. Structurally, the eigenvalues of $A$ can be used to identify community structures in the network~\cite{Chauhan2009PRE} and measure the large-scale connectivity of a graph~\cite{Taylor2012PRE}. The spectrum of the adjacency matrix also determines critical transition points in dynamical processes ranging from branching processes~\cite{Larremore2011PRL} and epidemic spreading~\cite{Pastor2001PRL} to synchronization~\cite{Restrepo2005PRE}.

In this work we study the eigenvalue spectrum of the adjacency matrix of the hierarchical product of two graphs along with the contribution from each of the eigenvalue spectrums of the underlying graphs. Using a combination of exact analytical results and perturbation theory, we derive analytical approximations for the full spectrum of eigenvalues of the hierarchical product as a function of the eigenvalues and eigenvectors of the two underlying networks and a coupling parameter that tunes their interactions. Due to the central role of the Perron-Frobenius (PF) eigenvalue~\cite{MacCluer2000SIAM}, i.e., the largest eigenvalue, of the adjacency matrix in several applications~\cite{Restrepo2006PRL,Restrepo2007PRE}, we study in detail the behavior of the PF eigenvalue. We observe that the PF eigenvalue tends to be minimized roughly when the coupling parameter is tuned to equally balance the contribution of the two underlying graphs -- a result that is supported by our analysis. Moreover, we investigate the role of the root set in connecting the hierarchical product and its impact on the PF eigenvalue. Finally, as an example application we consider the Susceptible-Infected-Susceptible (SIS) epidemic spreading model~\cite{Gomez2010EPL} on the hierarchical product of two graphs and use our theory to generate accurate predictions for the epidemic threshold between persistent infection and extinction of the epidemic.

The remainder of this paper is organized as follows. In Sec.~\ref{sec2} we define the hierarchical product of two graphs and discuss the overall behavior of the eigenvalue spectrum of the adjacency matrix as a function of the coupling parameter. In Sec.~\ref{sec3} we present a perturbation theory for the approximation of the eigenvalues of the adjacency matrix of the hierarchical product corresponding to limits when the coupling parameter is both small and large. In Sec.~\ref{sec4} we investigate in detail the behavior of the PF eigenvalue and the effect of the root set in connecting the hierarchical product. In Sec.~\ref{sec5} we present an example application of our theory in epidemic spreading on a network. Finally, In Sec.~\ref{sec6} we conclude with a discussion of our results.

\section{The Hierarchical Product}\label{sec2}

\begin{figure}[t]
\centering
\includegraphics[width=0.95\linewidth]{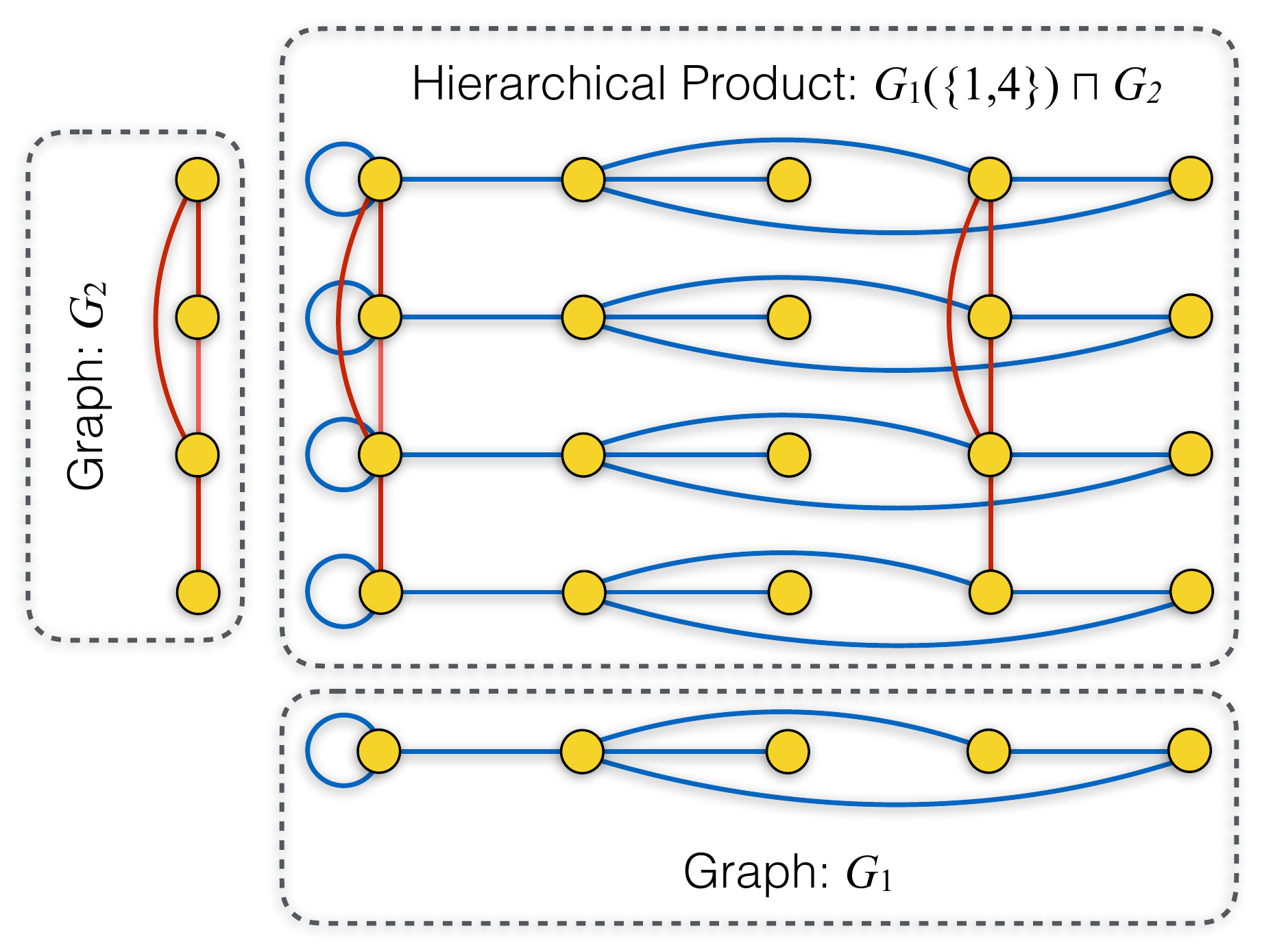}
\caption{(Color online) Illustration of the hierarchical product $G$ of two subgraphs $G_1$ and $G_2$ using the root set $U=\{1,4\}$.}
\label{fig1}
\end{figure}

\begin{figure*}[t]
\centering
\includegraphics[width=0.67\linewidth]{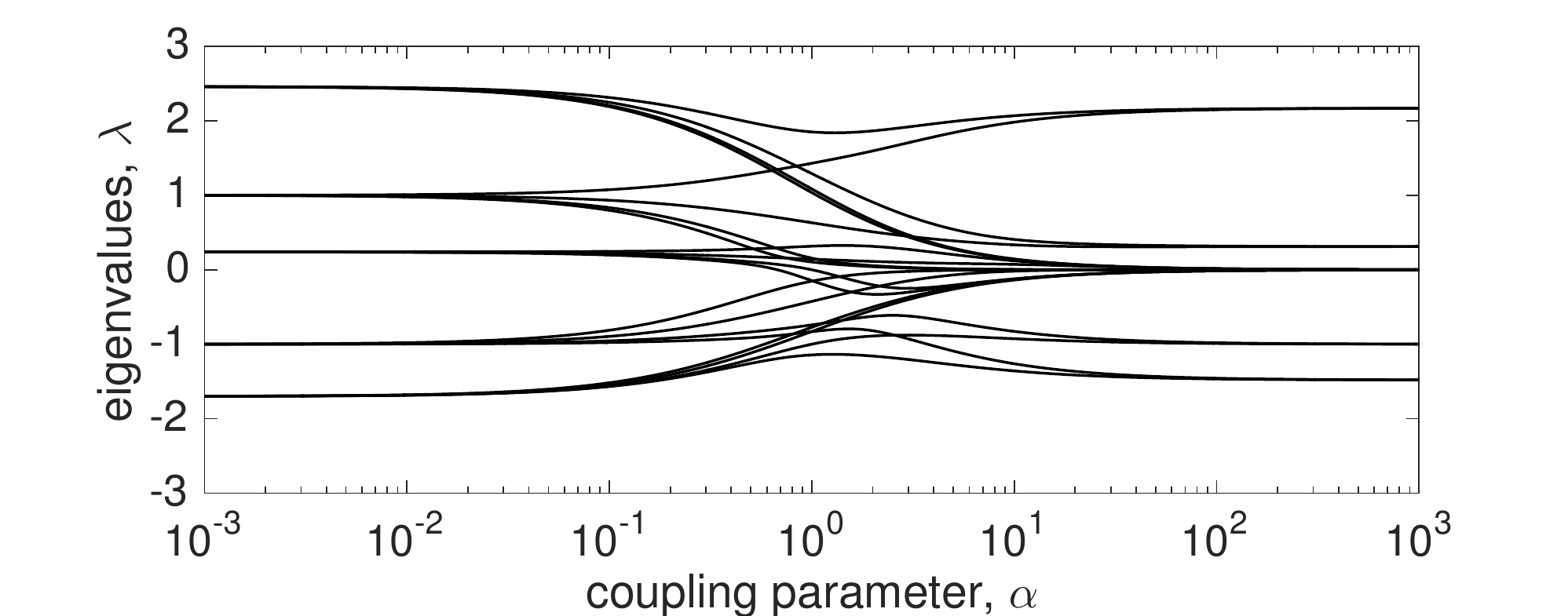}
\caption{(Color online) The eigenvalue spectrum of the hierarchical product defined in Fig.~\ref{fig1} as a function of the coupling parameter $\alpha$.}
\label{fig2}
\end{figure*}

The mathematical structure underlying any network is a graph $G$ consisting of a set $V(G)$ of vertices (sometimes called nodes) and a set $E(G)$ of edges (sometimes called links) connecting the vertices. We consider here the hierarchical product of two graphs, $G_1$ and $G_2$, defined as follows.
\begin{definition}
Given graphs $G_1$ and $G_2$ and any subset $U$ of vertices in $G_1$ referred to as the {\it root set}, the {\it hierarchical product}, denoted $G_1(U) \sqcap G_2$, is the graph $G$ with vertex set $V(G)=V(G_1)  \times V(G_2)$ whereby any two vertices $(x_1, y_1)$ and $(x_2, y_2)$ of $V(G)$ are adjacent if either $y_1 = y_2$ and $x_1x_2 \in E(G_1)$ or $x_1 = x_2$, $x_1 \in U$, and $y_1y_2 \in E(G_2)$.
\end{definition}
Letting $N_1$ and $N_2$ denote the order of the graphs $G_1$ and $G_2$, respectively, then the hierarchical product $G_1(U)\sqcap G_2$ is of order $N=N_1\cdot N_2$. The hierarchical product can thus be understood intuitively as $N_2$ copies of the graph $G_1$, which are themselves connected at the vertices included in the root set $U$ via the graph $G_2$. In Fig.~\ref{fig1} we illustrate the generic structure of a hierarchical product with an illustrative example of the roles of two graphs $G_1$ and $G_2$ of order $N_1=5$ and $N_2=4$, respectively, and root set $U=\{1,4\}$ in $G_1$. We note that the hierarchical product can be further generalized to include the product of an arbitrary number of graphs~\cite{Barriere2009DAM2}. However, as such hierarchical products can be defined recursively, we focus on hierarchical products of two graphs. 

Next, we introduce a coupling parameter to weigh the contributions of the graphs $G_1$ and $G_2$ to the hierarchical product $G=G_1(U)\sqcap G_2$. Denoting the coupling parameter $\alpha>0$, we weigh the links $G$ owing to $G_1$ and $G_2$ by the sigmoidal functions $(1+\alpha)^{-1}$ and $\alpha(1+\alpha)^{-1}$, respectively. Thus for $\alpha<1$ the graph $G_1$ is weighted more heavily than $G_2$, for $\alpha>1$ the graph $G_2$ is weighted more heavily than $G_1$, and for $\alpha=1$ the graphs $G_1$ and $G_2$ are weighted equally. To express the adjacency matrix of the hierarchical product we utilize the Kronecker product. Specifically, denoting the adjacency matrix with coupling $\alpha$ as $A_\alpha$, we have that
 \begin{align}
A_\alpha &= \left(I_2 \otimes A_1 + \alpha A_2 \otimes D_1\right)/(1+\alpha),\label{eq:adj01}%\\
%&\cong \left( A^1 \otimes I^2 + \alpha D^1 \otimes A^2\right)/(1+\alpha),\label{eq:adj02}
\end{align}
where $A_1$ and $A_2$ are the adjacency matrices associated to graphs $G_1$ and $G_2$, $I_2$ is the $N_2\times N_2$ identity matrix, and $D_1$ is the $N_1\times N_1$ diagonal matrix whose $i^{\text{th}}$ diagonal entry is equal to one if vertex $i$ is in the root set $U$ and zero otherwise otherwise. Thus, $D_1$ encodes the connections between the graphs $G_1$ and $G_2$ as defined by the root set $U$. For simplicity we focus on the case where $G_1$ and $G_2$ are both undirected and unweighted, and thus $A_1$ and $A_2$ are symmetric and binary, however these assumptions can be easily relaxed to generalize the results presented below.

Before proceeding to the analysis we use the example in Fig.~\ref{fig1} to investigate the generic behavior of the eigenvalue spectrum of the hierarchical product. In Fig.~\ref{fig2} we plot the eigenvalues of the adjacency matrix $A_\alpha$ as a function of the coupling parameter $\alpha$ (computed numerically). First, we note that the eigenvalues vary smoothly as a function of $\alpha$. Second, there are two different limiting behaviors as $\alpha$ is made very small and very large, respectively, with a complicated entanglement of eigenvalues in between where $\alpha$ is roughly of order one. In this particular example these limiting behaviors each consist of five limiting values to which all the eigenvalues approach, but in general the number of limiting values for small and large $\alpha$ need not match. Third, focusing our attention on the PF, or largest, eigenvalue, we observe that it attains a global minimum when $\alpha$ is of order one, i.e., when the contribution of $G_1$ and $G_2$ are roughly balanced. In the remainder of this paper we will present an asymptotic analysis for the behavior of the full spectrum of eigenvalues in the limits of both large and small $\alpha$ which will recover the exact limiting values of each eigenvalues well as the leading-order relaxation to these values. Moreover, our asymptotic analysis predicts the dip we observe in the PF eigenvalue, and can be used to accurately predict dynamical behavior on hierarchical products.

\section{Asymptotic Analysis}\label{sec3}
Our asymptotic analysis of the eigenvalue spectrum of the adjacency matrix $A_\alpha$ stems from an exact result for the eigenvalues spectrum of any matrix of the form in Eq.~(\ref{eq:adj01}). In particular, we have the following:
\begin{theorem}\cite{Barriere2009DAM1} \label{theorem}
Let $\{\mu_i\}_{i=1}^{N_2}$ be the collection of $N_2$ eigenvalues of $A_2$, and define
\begin{align}
A_\alpha(\mu_i) = (A_1+\alpha \mu_i D_1)/(1+\alpha)\label{eq:Matrix}
\end{align}
for each $i=1,\dots,N_2$. Then $\lambda$ is an eigenvalue of $A_\alpha$ as defined in Eq.~(\ref{eq:adj01}) if and only if it is an eigenvalue of $A_\alpha(\mu_i)$ for some $i=1,\dots,N_2$.
\end{theorem}
In particular, Theorem~\ref{theorem} expresses the eigenvalues of $A_\alpha$ as the collection of eigenvalues of each smaller matrix $A_\alpha(\mu_i)$. Since $N_2$ such smaller matrices exist, each with $N_1$ eigenvalues, we thus recover the full spectrum of $N=N_1\cdot N_2$ eigenvalues of the original adjacency matrix. 

Next we  perform the asymptotic analysis for the eigenvalues of $A_\alpha$ via the collection of matrices $A_\alpha(\mu_i)$, first in the limit of small $\alpha$, then in the limit of large $\alpha$. In particular, we will show that in both cases the full spectrum is determined by the eigenvalues and eigenvectors of $A_1$ and $A_2$, the entries of $D_1$, and the parameter $\alpha$. In the analysis below we will denote the eigenvalues and eigenvectors of $A_1$ as $\{\nu_i\}_{i=1}^{N_1}$ and $\{\bm{v}^i\}_{i=1}^{N_1}$, respectively, and the eigenvalues and eigenvectors of $A_2$ as $\{\mu_i\}_{i=1}^{N_2}$ and $\{\bm{u}^i\}_{i=1}^{N_2}$, respectively. Moreover, since $A_1$ and $A_2$ are both symmetric the sets of eigenvectors $\{\bm{v}^i\}_{i=1}^{N_1}$ and $\{\bm{u}^i\}_{i=1}^{N_2}$ can be appropriately normalized to form orthonormal bases for $\mathbb{R}^{N_1}$ and $\mathbb{R}^{N_2}$~\cite{Golub}, respectively such that $\bm{v}^{iT}\bm{v}^j=\delta_{ij}$ and $\bm{u}^{iT}\bm{u}^j=\delta_{ij}$.

\subsection{Perturbation Theory: Small $\alpha$}
We begin by considering the case where the coupling parameter $\alpha$ is small. Proceeding perturbatively, we let $\epsilon=\alpha$ such that $\epsilon\ll1$ is a small parameter and let 
\begin{align}
\tilde{A}_\epsilon(\mu_i) = A_1+\epsilon\mu_i D_1,\label{eq:Theory01}
\end{align}
such that $A_\epsilon(\mu_i)=(1+\epsilon)^{-1}\tilde{A}_\epsilon(\mu_i)$. We then search for the eigenvalues of the matrix $\tilde{A}_\epsilon(\mu_i)$ since its eigenvalues, denoted $\{\tilde{\lambda}_j(\epsilon)\}_{j=1}^{N_1}$, can be scaled by $(1+\epsilon)^{-1}$ to obtain the eigenvalues of $A_\epsilon(\mu_i)$, denoted $\{\lambda_j(\epsilon)\}_{j=1}^{N_1}$. We also denote the eigenvectors [of both $A_\epsilon(\mu_i)$ and $\tilde{A}_\epsilon(\mu_i)$] as $\{\bm{w}^j(\epsilon)\}_{j=1}^{N_1}$. In the limit $\epsilon\to0^+$, it is clear to see that the spectrum of $\tilde{A}_\alpha(\mu_i)$ is simply that of $A_1$, i.e., $\tilde{\lambda}_j(0)=\nu_i$ and $\bm{w}^j(0)=\bm{v}^j$. For $0<\epsilon\ll1$, we then propose the following perturbative ansatz for the eigenvalues and eigenvectors:
\begin{align}
\tilde{\lambda}_j(\epsilon)&=\nu_j+\epsilon\hat{\lambda}_j + \mathcal{O}(\epsilon^2),\label{eq:Theory02}\\
\bm{w}^j(\epsilon)&=\bm{v}^j+\epsilon\hat{\bm{w}}^j + \mathcal{O}(\epsilon^2).\label{eq:Theory03}
\end{align}
Inserting Eqs.~(\ref{eq:Theory01}), (\ref{eq:Theory02}), and (\ref{eq:Theory03}) into the eigenvalue equation $\tilde{A}_\epsilon(\mu_i)\bm{w}^j(\epsilon)=\tilde{\lambda}_j(\epsilon)\bm{w}^j(\epsilon)$ and collecting the leading order terms at $\mathcal{O}(\epsilon)$, we obtain
\begin{align}
\mu_i D_1\bm{v}^j + A_1 \hat{\bm{w}}^j=\hat{\lambda}_j\bm{v}^j+\nu_j\hat{\bm{w}}^j.\label{eq:Theory04}
\end{align}
Left-multiplying Eq.~(\ref{eq:Theory04}) by $\bm{v}^{jT}$ and noting that the term on the left-hand side $\bm{v}^jA_1\hat{\bm{w}}^j=\nu_j\bm{v}^j\hat{\bm{w}}^j$ cancels with the right-hand side, we obtain
\begin{align}
\hat{\lambda}_j=\mu_i\bm{v}^{jT}D_1\bm{v}^j.\label{eq:Theory05}
\end{align}
Multiplying by $(1+\epsilon)^{-1}$ to recover $\lambda(\epsilon)$ and substituting back $\epsilon=\alpha$, we have that the eigenvalues of $A_\alpha(\mu_i)$ to leading order are given by
\begin{align}
\lambda_j(\alpha)=\frac{\nu_j+\alpha\mu_i\bm{v}^{jT}D_1\bm{v}^j}{1+\alpha}.\label{eq:Theory06}
\end{align}
The full spectrum of eigenvalues of $A_\alpha$ is then the collection of all eigenvalues $\lambda_j(\alpha)$ for $j=1,\dots,N_1$ given in Eq.~(\ref{eq:Theory06}) evaluated at each $\mu_i$ for $i=1,\dots,N_2$.

\begin{figure}[b]
\centering
\includegraphics[width=0.95\linewidth]{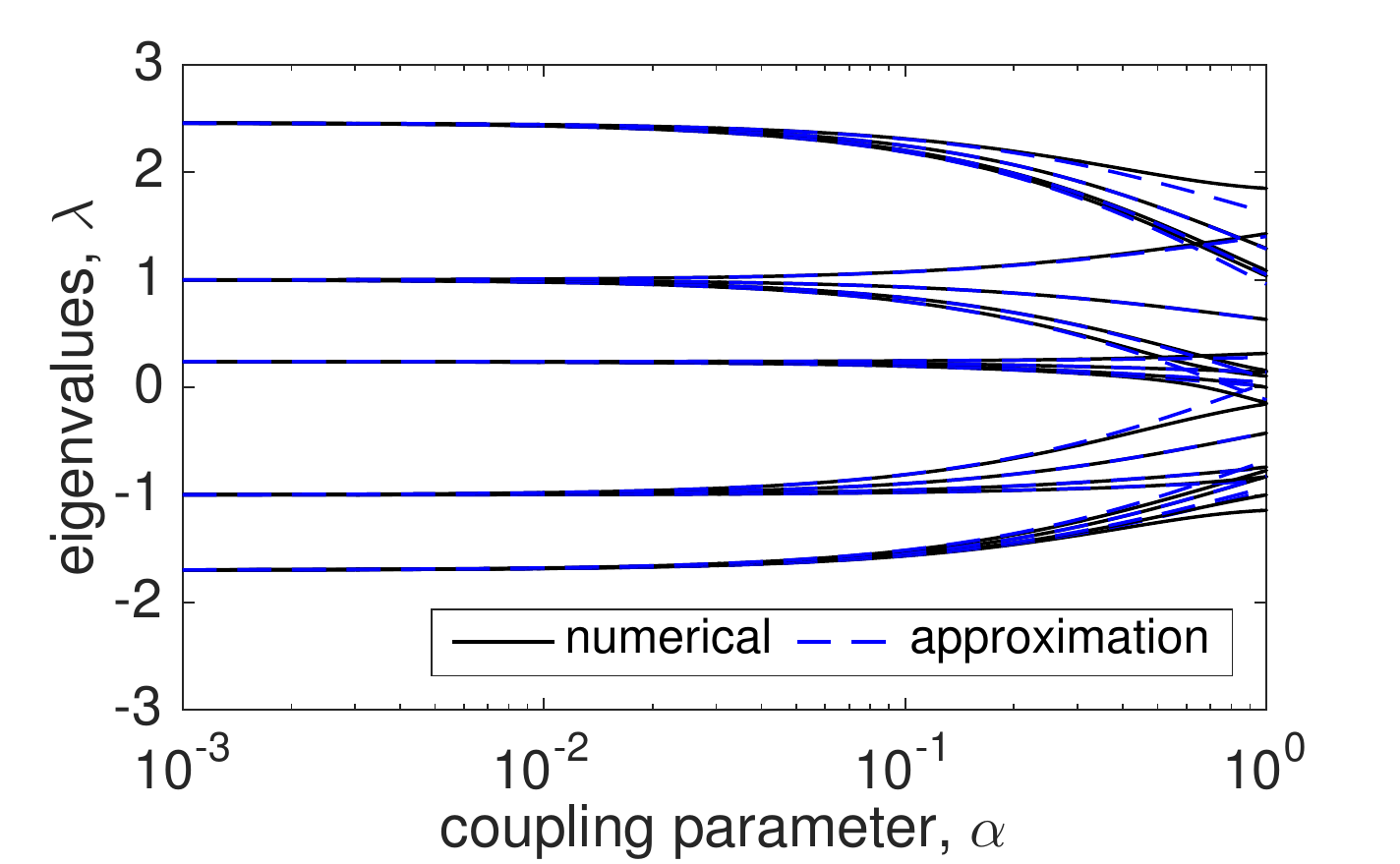}
\caption{(Color online) {\it Asymptotic approximation: small $\alpha$}. Asymptotic approximation [Eq.~(\ref{eq:Theory06})] (dashed blue) vs numerically calculated eigenvalues (solid black) of the adjacency matrix for the hierarchical product illustrated in Fig.~\ref{fig1} for small values of the coupling parameter $\alpha$.}
\label{fig3}
\end{figure}

By considering the limit $\alpha\to0^+$ of Eq.~(\ref{eq:Theory06}) we recover the exact limiting values of the eigenvalues of $A_\alpha$ for small $\alpha$. In particular, in this limit we have that $\lambda_j(0)=\nu_j$, and therefore the spectrum of $A_\alpha$ contains $N_2$ copies each of the eigenvalues $\nu_j$ of $A_1$. Moreover, the first-order correction describing the relaxation towards these limiting values is determined by the term $\mu_i\bm{v}^{jT}D_1\bm{v}^j$, i.e., the action of $D_1$ on the $j^{\text{th}}$ eigenvector of $A_1$ scaled by the eigenvalues of $A_2$. Using the example illustrated in Fig.~\ref{fig1}, we compare our asymptotic approximation for the eigenvalues of the adjacency matrix $A_\alpha$ to its actual eigenvalues, plotting in Fig.~\ref{fig3} the approximation (dashed blue) and the numerically calculated eigenvalues (solid black) for $\alpha<1$. We observe a strong agreement between the numerical and approximate eigenvalues  $\alpha$, which only loses accuracy when $\alpha$ becomes roughly order one, where the asymptotic analysis is expected to break down.

\subsection{Perturbation Theory: Large $\alpha$}
Next we consider the case where the coupling parameter $\alpha$ is large. We again proceed perturbatively, now letting $\epsilon=\alpha^{-1}$ such that $\epsilon\ll1$ is a small parameter and let
\begin{align}
\tilde{A}_\epsilon(\mu_i)=\mu_i D_1 + \epsilon A_1,\label{eq:Theory07}
\end{align}
such that $A_\epsilon(\mu_i)=\epsilon^{-1}(1+\epsilon^{-1})^{-1}\tilde{A}_\epsilon(\mu_i)$. Similarly, we search for the eigenvalues $\tilde{\lambda}_j(\epsilon)$ of $\tilde{A}_\epsilon(\mu_i)$ which we use to recover the eigenvalues of $A_\epsilon(\mu_i)$ after scaling by $\epsilon^{-1}(1+\epsilon^{-1})^{-1}$. We first point out that for $\epsilon=0$ the matrix $\tilde{A}_0(\mu_i)$ reduces to $\mu_iD_1$, which is highly degenerate. Specifically, if the root set $U$ contains $n$ connecting vertices, then $D_1$ has $n$ eigenvalues equal to one and $(N_1-n)$ eigenvalues equal to zero. Moreover, the nontrivial eigenspace of $D_1$ is precisely the span of all vectors whose entries are zero where the diagonal entries of $D_1$ are zero, and the trivial eigenspace (i.e., the nullspace) of $D_1$ is precisely the span of all vectors whose entries are zero where the diagonal entries of $D_1$ are non-zero. Due to this dichotomy, the asymptotic analysis of the eigenvalues of $\tilde{A}_\epsilon(\mu_i)$ splits into two cases: one for the $n$ eigenvalues associated with the non-trivial eigenspace of $D_1$ and another for the $(N_1-n)$ eigenvalues associated with the nullspace of $D_1$.

We begin with the non-trivial eigenspace of $D_1$, proposing the perturbative ansatz
\begin{align}
\tilde{\lambda}_j(\epsilon)&=\mu_i+\epsilon\hat{\lambda}_j + \mathcal{O}(\epsilon^2),\label{eq:Theory09}\\
\bm{w}^j(\epsilon)&=\bm{x}+\epsilon\hat{\bm{w}}^j + \mathcal{O}(\epsilon^2),\label{eq:Theory10}
\end{align}
where the vector $\bm{x}$ is in the non-trivial nullspace of $D_1$, i.e., $D_1\bm{x}=\bm{x}$. Inserting Eqs.~(\ref{eq:Theory07}), (\ref{eq:Theory09}), and (\ref{eq:Theory10}) into the eigenvalue equation $\tilde{A}_\epsilon(\mu_i)\bm{w}^j(\epsilon)=\tilde{\lambda}_j(\epsilon)\bm{w}^j(\epsilon)$ and collecting the leading order terms at $\mathcal{O}(\epsilon)$, we obtain
\begin{align}
\mu_i D_1\hat{\bm{w}}^j+A_1\bm{x}=\hat{\lambda}_j\bm{x}+\mu_i\hat{\bm{w}}^j.\label{eq:Theory11}
\end{align}
Inspecting Eq.~(\ref{eq:Theory11}) and noting that for each diagonal entry of $D_1$ that is zero, the corresponding entry of the left-hand side is zero, we find that so must the corresponding entries on the right-hand side. Eliminating these entries, we obtain the $n$-dimensional vector equation
\begin{align}
\mu_i\hat{\bm{w}}^{\cancel{0}}+A_1^{\cancel{0}}\bm{x}^{\cancel{0}}&=\hat{\lambda}_j\bm{x}^{\cancel{0}}+\mu_i\hat{\bm{w}}^{\cancel{0}},\\
\to\hskip8exA_1^{\cancel{0}}\bm{x}^{\cancel{0}}&=\hat{\lambda}_j\bm{x}^{\cancel{0}},\label{eq:Theory12}
\end{align}
where $A_1^{\cancel{0}}$ is the $n\times n$ matrix obtained by keeping the rows and columns of $A_1$ corresponding to non-zero diagonal entries of $D_1$ and similarly $\hat{\bm{w}}^{j\cancel{0}}$ and $\bm{x}^{\cancel{0}}$ are the $n$-dimensional vectors obtained by keeping the entries of $\hat{\bm{w}}^j$ and $\bm{x}$ corresponding to non-zero entries of $D_1$. Thus, $\hat{\lambda}_j$ is given by the $j^{\text{th}}$ eigenvalue of the matrix $A_1^{\cancel{0}}$, denoted $\nu_j^{\cancel{0}}$. Thus, the $n$ eigenvalues of $A_\alpha(\mu_i)$ corresponding to the non-zero eigenspace of $D_1$ to leading order are given by
\begin{align}
\lambda_j(\alpha)=\frac{\alpha\mu_i+\nu_j^{\cancel{0}}}{1+\alpha},\label{eq:Theory13}
\end{align}
which approaches the value $\mu_i$ in the limit $\alpha\to\infty$.

\begin{figure}[t]
\centering
\includegraphics[width=0.95\linewidth]{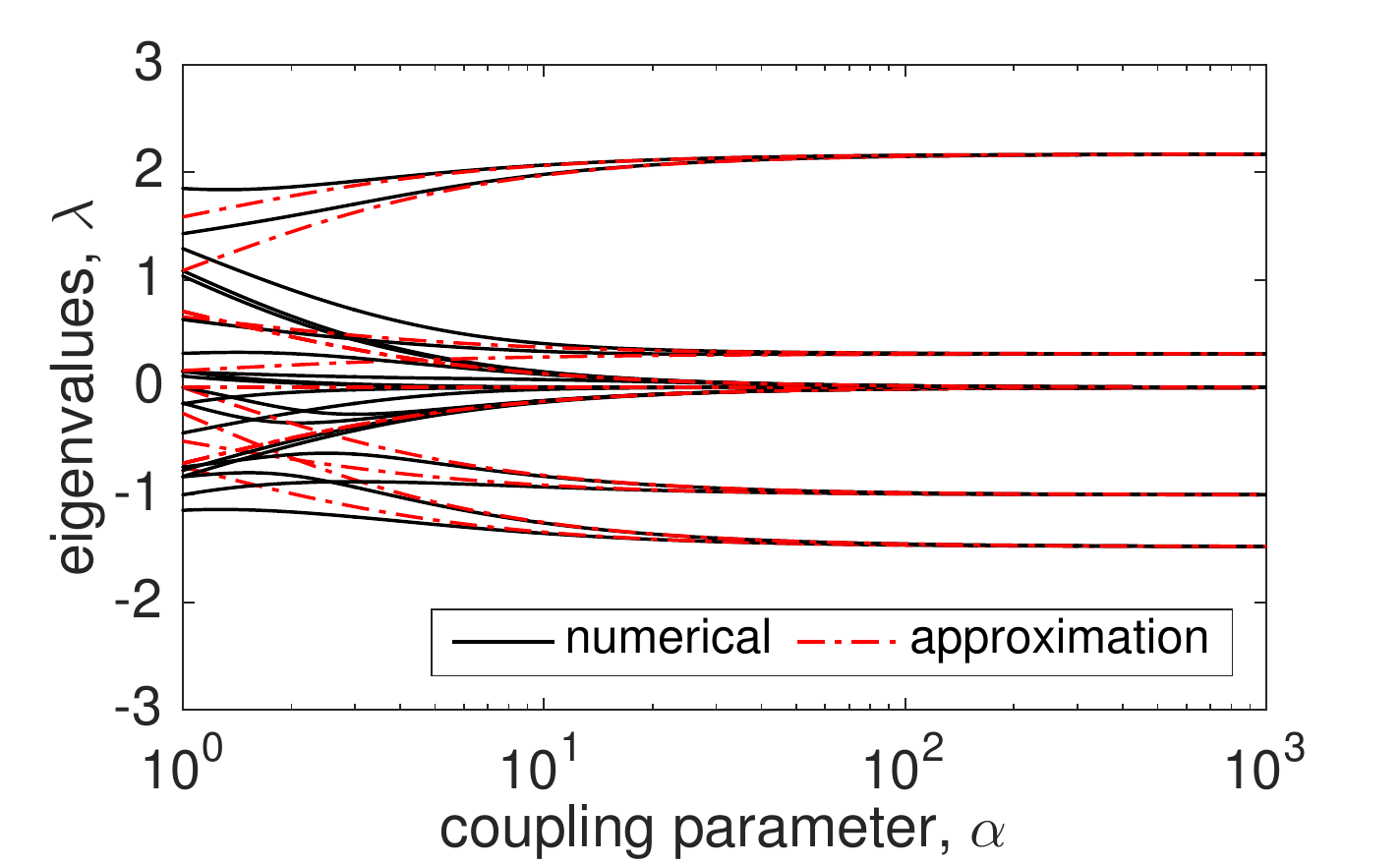}
\caption{(Color online) {\it Asymptotic approximation: large $\alpha$}. Asymptotic approximation [Eq.~(\ref{eq:Theory13}) and (\ref{eq:Theory18})] (dot-dashed red) vs numerically calculated eigenvalues (solid black) of the adjacency matrix for the hierarchical product illustrated in Fig.~\ref{fig1} for large values of the coupling parameter $\alpha$.}
\label{fig4}
\end{figure}

For the remaining $(N_1-n)$ eigenvalues of $\tilde{A}_\epsilon(\mu_i)$ associated with the nullspace of $D_1$, we introduce a different perturbative anstaz:
\begin{align}
\tilde{\lambda}_j(\epsilon)&=0+\epsilon\hat{\lambda}_j + \mathcal{O}(\epsilon^2),\label{eq:Theory14}\\
\bm{w}^j(\epsilon)&=\bm{x}+\epsilon\hat{\bm{w}}^j + \mathcal{O}(\epsilon^2),\label{eq:Theory15}
\end{align}
where the vector $\bm{x}$ is now in the nullspace of $D_1$, i.e., $D_1\bm{x}=\bm{0}$. Inserting Eqs.~(\ref{eq:Theory07}), (\ref{eq:Theory14}), and (\ref{eq:Theory15}) into the eigenvalue equation $\tilde{A}_\epsilon(\mu_i)\bm{w}^j(\epsilon)=\tilde{\lambda}_j(\epsilon)\bm{w}^j(\epsilon)$ and collecting the leading order terms at $\mathcal{O}(\epsilon)$, we obtain
\begin{align}
\mu_i D_1\hat{\bm{w}}^j+A_1\bm{x}=\hat{\lambda}_j\bm{x}.\label{eq:Theory16}
\end{align}
Inspecting Eq.~(\ref{eq:Theory16}) and noting for each non-zero diagonal entry of $D_1$ the corresponding entry of $\bm{x}$ is zero, we eliminate each of these entries and find $\bm{x}$ corresponding to non-zero diagonal entries of $D_1$ is itself zero. Eliminating these entries, we obtain the $(N_1-n)$-dimensional vector equation
\begin{align}
A_1^0\bm{x}^0=\hat{\lambda}_j\bm{x}^0,\label{eq:Theory17}
\end{align}
where $A_1^0$ is the $(N_1-n)\times(N_1-n)$ matrix obtained by keeping the rows and columns of $A_1$ corresponding to zero diagonal entries of $D_1$ and similarly $\bm{x}^0$ is the $(N_1-n)$-dimensional vector obtained by keeping the entries of $\bm{x}$ corresponding to zero entries of $D_1$. Thus, $\hat{\lambda}_j$ is given by the $j^{\text{th}}$ eigenvalue of the matrix $A_1^0$, denoted $\nu_j^0$ and the $(N_1-n)$ eigenvalues of $A_\alpha(\mu_i)$ corresponding to the nullspace of $D_1$ to leading order are given by
\begin{align}
\lambda_j(\alpha)=\frac{\nu_j^0}{1+\alpha},\label{eq:Theory18}
\end{align}
all of which approach zero as $\alpha\to\infty$.

Combining the asymptotic analysis for the spectrum of $A_\alpha(\mu_i)$ stemming from both the nontrivial eigenspace of $D_1$ and the nullspace of $D_1$, we obtain for each $\mu_i$ a collection of $n$ eigenvalues of the form in Eq.~(\ref{eq:Theory13}) along with $(N_1-n)$ eigenvalues of the form in Eq.~(\ref{eq:Theory18}). Moreover, in the limit of large $\alpha$ eigenvalues of the form in Eq.~(\ref{eq:Theory13}) each approach the limiting value $\mu_i$ while eigenvalues of the form in Eq.~(\ref{eq:Theory18}) each approach a limiting value of zero, while the relaxation to these values are determined by the eigenvalues of the matrices $A_1^{\cancel{0}}$ and $A_1^0$, respectively. Thus, assuming that each eigenvalue $\mu_i$ of $A_2$ is distinct, in total the spectrum of $A_\alpha$ will have $n$ eigenvalues each that limit to each distinct $\mu_i$ and $N_2\left(N_1-n\right)$ eigenvalues that limit to zero. Again using the example illustrated in Fig.~\ref{fig1}, we compare our asymptotic approximation for the eigenvalues of the adjacency matrix $A_\alpha$ to its actual eigenvalues, plotting in Fig~\ref{fig4} the the approximation (dot-dashed red) and the numerically calculated eigenvalues (solid black) for $\alpha>1$. We observe a strong agreement between the numerical and approximate eigenvalues  $\alpha$, which only loses accuracy when $\alpha$ becomes roughly order one, where the asymptotic analysis is expected to break down.

\section{Perron-Frobenius Eigenvalue}\label{sec4}
The Perron-Frobenius theorem guarantees that for any network with non-negative and irreducible adjacency matrix $A$ the eigenvalue with largest magnitude is real, positive, and distinct. We call this largest eigenvalue the Perron-Frobenius eigenvalue~\cite{MacCluer2000SIAM} and denote it
\begin{align}
\Lambda=\sup_{\lambda_i\in\sigma(A)}\lambda_i,\label{eq:PF}
\end{align}
where $\sigma(A)$ denotes the eigenvalue spectrum of $A$. In a wide range of dynamical processes on networks the PF eigenvalue plays an especially important role in shaping the macroscopic steady-state behavior~\cite{Restrepo2007PRE}. For instance, in the case of the SIS epidemic model the critical infection rate delineating the persistence or extinction of the epidemic is proportional to the inverse of the PF eigenvalue~\cite{Gomez2010EPL}. Another example lies in the synchronization of large networks of coupled oscillators, where the critical coupling strength corresponding to the onset of synchronization is also proportional to the inverse of the PF eigenvalue~\cite{Restrepo2005PRE}. Thus, in many cases the PF eigenvalue can be used as a quantitative measure for the connectivity of a network~\cite{Taylor2012PRE}. Given its importance, we now focus our attention on the PF eigenvalue of hierarchical products.

\begin{figure}[t]
\centering
\includegraphics[width=0.95\linewidth]{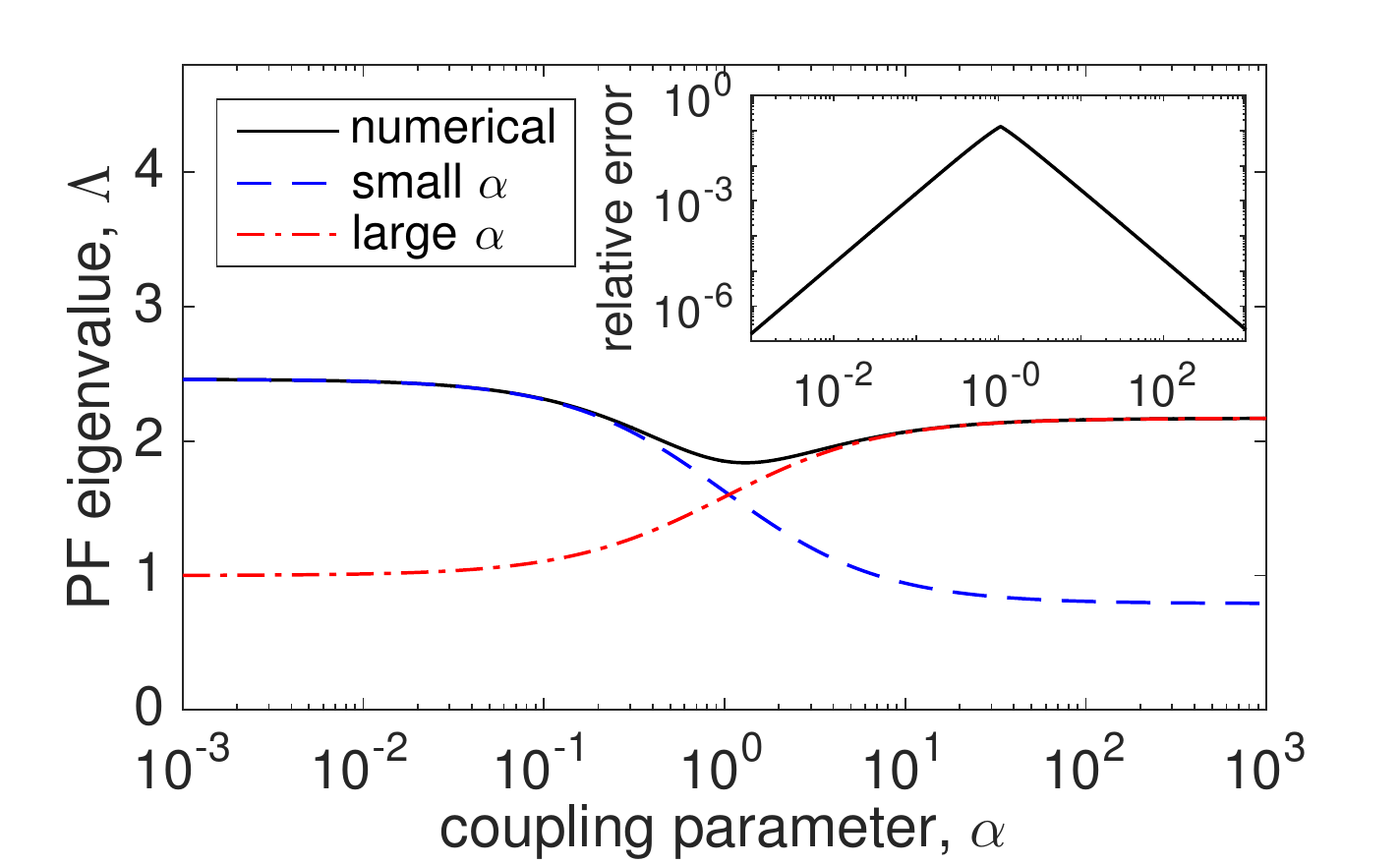}
\caption{(Color online) {\it PF eigenvalue: asymptotic approximations}. For the hierarchical product illustrated in Fig.~\ref{fig1}, the PF eigenvalue calculated numerically (solid black) and given by the asymptotic approximations for both small and large $\alpha$ in Eqs.~(\ref{eq:Theory06}) and (\ref{eq:Theory13}) (dashed blue and dot-dashed red, respectively) as a function of $\alpha$. Inset: relative error.}
\label{fig5}
\end{figure}

In the respective limits of small and large $\alpha$, the asymptotic approximations for the PF eigenvalue are given by Eqs.~(\ref{eq:Theory06}) and (\ref{eq:Theory13}), using the largest eigenvalues of $A_1$, $A_2$, and $A_1^{\cancel{0}}$, i.e., $\nu_{\text{max}}$, $\mu_{\text{max}}$, and $\nu_{\text{max}}^{\cancel{0}}$. Using the example illustrated in Fig.~\ref{fig1} we plot in Fig.~\ref{fig5} the PF eigenvalue of $A_\alpha$ calculated numerically (solid black) as well as the approximations for small and large $\alpha$ (dashed blu and dot-dashed red, respectively). Taking the overall asymptotic approximation as the maximum of the two approximations for small and large $\alpha$, we also plot the relative error of our approximation in the inset. Similar to the results for the full eigenvalue spectrum, the asymptotic approximations holds very well, breaking down only when $\alpha$ is roughly of order one. Moreover, we observe that as $\alpha$ approaches the order one regime, the approximations for both small and large $\alpha$ in fact decrease, guiding the PF eigenvalue to its dip near $\alpha\approx1$ as was originally observed.

In addition to the overall behavior of the PF eigenvalue, we also consider the effect of different root sets $U$ that define the hierarchical product $G_1(U)\sqcap G_2$. Recall that the vertices in $U$ correspond to the non-zero entries of the matrix $D_1$ in Eq.~(\ref{eq:adj01}). What then is the result of using different root sets in generating the hierarchical product of two graphs? In particular, how does the PF eigenvalue behave depending on whether the root set is made up of well-connected or poorly-connected vertices? 

\begin{figure}[t]
\centering
\includegraphics[width=0.95\linewidth]{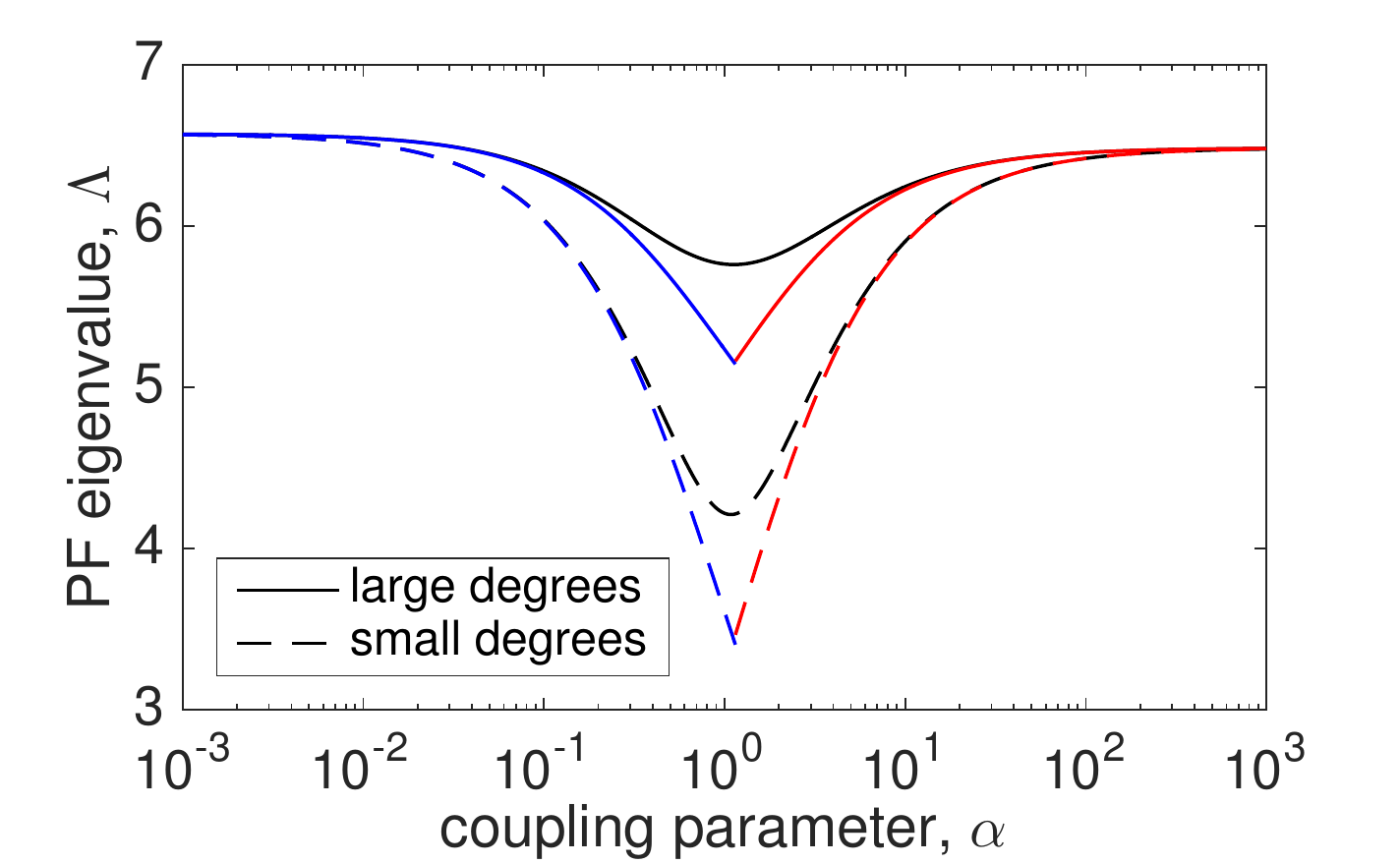}
\caption{(Color online) {\it Effect of root sets}. The numerically calculated PF eigenvalue for the hierarchical product of two BA graphs of size $N=20$ with minimum degree $k_0=3$ with root sets consisting of the five vertices in $G_1$ with largest degrees (solid black) and the five vertices in $G_2$ with smallest degrees (dashed black). Asymptotic approximations for both cases are plotted in blue and red.}
\label{fig6}
\end{figure}

We address this question by studying hierarchical products constructed from larger graphs generated by the Barabas\'{i}-Albert (BA) model~\cite{Barabasi1999Science}. In particular, the BA model is known for generating graphs with scale-free degree distributions and emerging hubs -- a relatively small number of vertices with many edges amid a majority of vertices with only a handful of edges. Thus, the BA model allows us the possibility to choose connecting sets made up of either well-connected or poorly-connected vertices. As an illustrative example we consider the hierarchical products of two BA graphs $G_1$ and $G_2$ both of size $N=20$ with minimum degree $k_0=3$. Using root sets $U$ of $n=5$ vertices, we create two distinct hierarchical products by choosing two different root sets: one consisting of the $n$ vertices with the largest degrees and another consisting of the $n$ vertices with the smallest degrees. In Fig.~\ref{fig6} we plot the numerically calculated PF eigenvalues of the hierarchical products built with the connecting sets of large degrees (solid black) and small degrees (dashed black), as well as the asymptotic approximations in blue and red. In particular, we observe that the dip in the PF eigenvalue is much more pronounced when the connecting set is made up of poorly-connected nodes. Thus, the connecting set made up of well-connected nodes preserves a much larger PF eigenvalue for all $\alpha$ values -- especially when $\alpha$ is roughly order one. However, we note that for both very large and very small $\alpha$ the choice of the root set has little effect on the PF eigenvalue.

\section{Application: Epidemic Spreading}\label{sec5}
As an application of our theory we now consider the SIS epidemic model on the hierarchical product of two graphs~\cite{Pastor2001PRL}. Given an underlying graph structure, the SIS model consists of two parameters: an infection rate $\beta$ and a healing rate $\beta$. Denoting the state of a node $i$ as $x_i=1$ if it is infected and $x_i=0$ if it is healthy the model evolves as follows. At each given time step $\Delta t\ll1$, each healthy node can itself be infected by any of its infected network neighbors $j$ with a probability of $\Delta t\beta A_{ij}$, while each infected node is healed and becomes healthy with probability $\Delta t \gamma$. Characterizing the macroscopic system state using the fraction of infected nodes, $X=N^{-1}\sum_{i=1}^Nx_i$ G\'{o}mez et al. showed in Ref.~\cite{Gomez2010EPL} that the critical epidemic threshold that delineates extinction of the epidemic, i.e., $X=0$, from long-time persistence of the epidemic, i.e., $X>0$, is given when the ratio of the infection rate to the healing rate is equal to the inverse of the PF eigenvalue, i.e.,
\begin{align}
\beta_c=\frac{\gamma}{\Lambda}.\label{eq:threshold}
\end{align}
In other words, if $\beta<\gamma/\Lambda$ the epidemic will eventually die out, and if $\beta>\gamma/\Lambda$ then the epidemic will persist for all time.

\begin{figure}[t]
\centering
\includegraphics[width=0.95\linewidth]{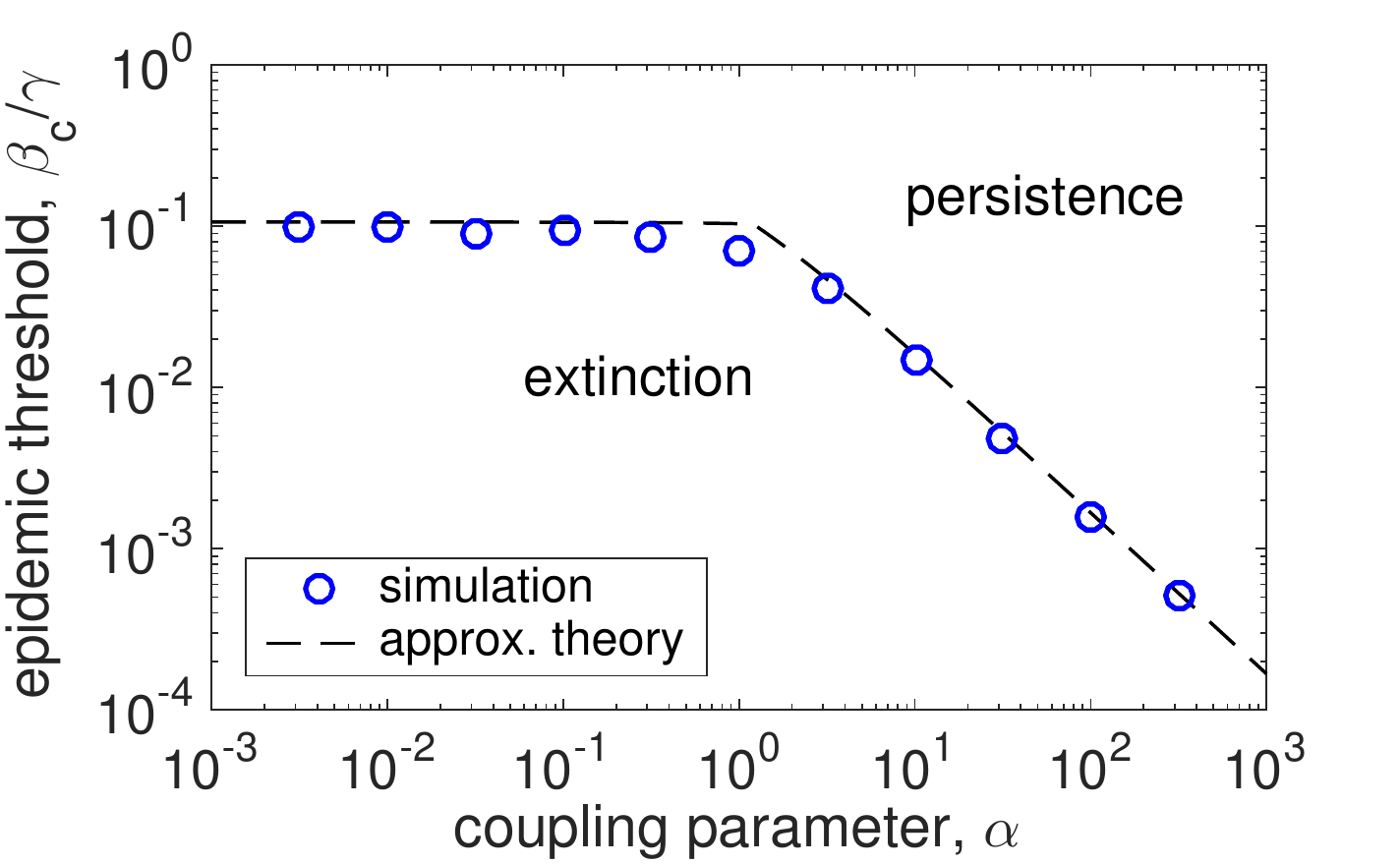}
\caption{(Color online) {\it Epidemic spreading}. Epidemic threshold $\beta_c/\gamma$ for the SIS model vs the coupling parameter $\alpha$ as computed directly from simulation (blue circles) and from our analytical predictions (dashed black) using the adjacency matrix in Eq.~(\ref{eq:adj02}). The underlying graph is a hierarchical product of a BA graph $G_1$ of size $N=100$ and a BA graph $G_2$ of size $N=20$, both with minimum degree $k_0=3$, and a connecting set of $n=20$ randomly chosen vertices in $G_1$.}
\label{fig7}
\end{figure}

To explore the behavior of the SIS model on a hierarchical product we consider a larger BA graph $G_1$ of size $N=100$ and minimum degree $k_0=3$ with a smaller BA graph $G_2$ of size $N=20$ and minimum degree $k_0=3$. We use a root set $U$ of $n=20$ randomly chosen vertices in $G_1$. Moreover, we take the larger graph $G_1$ to be fixed and scale the contribution of the smaller graph $G_2$ by the coupling parameter $\alpha$. The physical interpretation of this setup is to consider $G_1$ to be the primary, fixed graph while $G_2$ represents added transmission lines along which the epidemic spreads more slowly or quickly in comparison to $G_1$ depending on the value of $\alpha$. With this model setup we obtain a modified adjacency matrix,
\begin{align}
A_\alpha=I_2\otimes A_1+\alpha A_2\otimes D_1,\label{eq:adj02}
\end{align}
which is equivalent to that defined in Eq.~(\ref{eq:adj01}) after removing the factor $(1+\alpha)^{-1}$, and therefore its eigenvalues are also equivalent up to this rescaling. In Fig.~\ref{fig7} we present the results, plotting the epidemic threshold $\beta_c/\gamma$ (in our simulation we take $\gamma=1$) as observed from direct simulations of the model in blue circles vs the epidemic threshold as predicted from our asymptotic analysis of the PF eigenvalue in dashed black. Recall that any ratio $\beta/\gamma$ larger than the epidemic threshold leads to persistence of the epidemic, while any ratio smaller than the epidemic threshold leads to extinction of the epidemic. We note a strong agreement between the simulations and our analytical predictions, with the largest error near $\alpha\approx1$ as expected. We also observe a sharp transition in long-term behavior as a function of the coupling parameter. In particular, for $\alpha\lesssim1$ the epidemic threshold remains nearly constant, indicating that the graph $G_2$ contributes little to the overall spread of the epidemic. The transition then occurs at $\alpha\approx1$, after which the epidemic threshold decreases roughly as a power-law as $\alpha$ increases, indicating that the stronger contribution of $G_2$ allows for a quicker spread of the epidemic.

\section{Discussion}\label{sec6}

In this paper we have studied the spectral properties of the adjacency matrix of the hierarchical graph product of two smaller graphs. Using a blend of exact analytical results and an asymptotic analysis we have derived asymptotic approximations for the full spectrum of eigenvalues in the small and large limits of a coupling parameter introduced to weigh the relative contribution of each of the two smaller graphs. In particular, these asymptotic approximations are expressed in terms of the eigenvalues and eigenvectors of the two smaller graphs, simple properties of the roots set matrix, and the coupling parameter. These asymptotic approximations yield the exact limiting values of each eigenvalue in the limits when the coupling parameter is both small and large, as well as the first-order relaxation to these values.

Given its importance in dynamical phenomena including epidemic spreading and synchronization, we have studied in detail the behavior of the PF, or largest, eigenvalue. Interestingly, we observe that the PF eigenvalue reaches a global minimum when the two smaller graphs that make up the hierarchical product are roughly equally weighted, corresponding to when the coupling parameter is of order one. Although our asymptotic approximations are the least accurate in this regime, they do in fact predict this dip in the PF eigenvalue, decreasing as the coupling parameter approaches the order one regime. Moreover, we have investigated the effect of the choice of the root set on the PF eigenvalue. Specifically, when the root set is comprised of poorly-connected vertices this dip in the PF eigenvalue is accentuated, while when the root set is comprised of well-connected vertices this dip is less pronounced (albeit still present). Finally, as an application of our theory, we have studied the dynamics of the SIS epidemic model on hierarchical products, accurately predicting the epidemic threshold (i.e., the critical transition delineating the long-time persistence or extinction of the epidemic) as a function of the coupling parameter.

The construction of a graph via the hierarchical product represents a new method for building a large structure $G$ from smaller structures $G_1,\dots$~\cite{Barriere2009DAM1,Barriere2009DAM2}. Compared to the Cartesian product~\cite{Hammack2011} (of which the hierarchical product is a generalization) the hierarchical product captures connectivity characteristics that are less uniform and therefore promotes heterogeneity throughout the graph. Given the role of the eigenvalue spectra of various connectivity matrices in determining both dynamical and structural properties of the underlying graphs, the study of these eigenvalue spectra and their analytical approximations is an important direction of research. In this paper we have focused on the eigenvalue spectrum of the adjacency matrix, and in this way our work fits in the larger framework of studying the spectral properties of interconnected and multilayer networks to inform both the behavior of both linear and nonlinear dynamical behaviors~\cite{Sole2013PRE,Granell2013PRL,Hernandez2014PhysA,Kouvaris2015SciRep}. Important future work includes the eigenvalue spectra of other coupling matrices of hierarchical products. One such example with many physical applications is the eigenvalue spectrum of the combinatorial graph Laplacian, which plays an important role in shaping diffusion processes on graphs~\cite{Gomez2013PRL} as well as determining a networks' synchronization properties~\cite{Pecora1998PRL,Skardal2014PRL}. Another important coupling matrix is the modularity matrix which determines the community structures that make up a graph~\cite{Newman2006PRE}.

%\acknowledgements
%This work was funded in part by the James S. McDonnell Foundation (PSS and AA), NSF Grant No. DMS-1127914 through the Statistical and Applied Mathematical Sciences Institute (DT), Simons Foundation Grant No. 318812 (JS), Spanish DGICYT Grant No. FIS2012-38266 (AA), and FwET Project No. MULTIPLEX (317532) (AA).

\bibliographystyle{plain}

\end{document}